\documentclass[12pt]{iopart}
\usepackage[dvips]{graphicx}
\begin{document}

\article[Non-Markovian quantum trajectories]{ACOLS98}{A non-Markovian quantum trajectory approach to radiation into structured continuum}
\author{M. W. Jack\footnote{mwj@phy.auckland.ac.nz}, M. J. Collett and D. F. Walls}
\address{Department of Physics, University of Auckland, Private Bag 92019, New 
Zealand}

\begin{abstract}
We present a non-Markovian quantum trajectory method for treating atoms  radiating into a reservoir with a non-flat density of states.  The results of an example numerical simulation of the case where the free space modes of the reservoir are altered by the presence of a cavity are presented and compared with those of an extended system approach.
\end{abstract}
Submitted to: {\em J Opt. B: Special Issue - ACOLS 98}
\section{Introduction}

The radiative properties of an atom depend critically on the nature of the electromagnetic field surrounding the atom. It is therefore of interest to consider atoms situated in cavities \cite{purcell,kleppner,haroche89,raizen,berman}, wave guides and optical fibers \cite{burstein,brorson} or  photonic band gap materials \cite{john84,yablonovitch,kofman,john94,bay96} where the atoms have been shown to experience a very different process of decay from that in free space. In many of these cases the rapidly varying mode structure of the radiation reservoir invalidates the Born-Markov approximations usually employed to treat atoms radiating into free space \cite{bay97}. As a consequence of this problem, many theoretical works have considered these situations by treating the complete unitary dynamics of atom(s) and field \cite{kofman,john94,bay96}.

When the Born-Markov approximation applies, the reservoir degrees of freedom can be eliminated to derive a master equation for the reduced dynamics of the atom \cite{louisell}. Quantum trajectories  \cite{carmichael} (stochastic Schr\"{o}dinger equations \cite{gisin} or Monte Carlo wavefunctions \cite{dalibard}) have been found very successful in numerically solving this master equation \cite{gardiner,dum,tian,molmer}. A generalization of the quantum trajectory method beyond the Born-Markov regime has been made independently by Di\'{o}si {\it et al.} \cite{diosi98} and by the present authors \cite{jack}. This generalization opens the door to a treatment of atomic radiation into general reservoirs in terms of a reduced atomic system. By adhering to the measurement interpretation of quantum trajectories it is possible to bypass the intermediate step of deriving a master equation and go directly to a non-Markovian equation for the state of the system conditioned on a continuous measurement process \cite{jack}. Where the Born approximation does not hold this state is no longer pure owing to the presence of an entanglement between the reservoir and the atom.

In this paper we summarize the derivation of the non-Markovian trajectories and explicitly demonstrate how to simulate the trajectories for an atom radiating into the above mentioned structured continuum. The simulation method is viable if the number of time steps (where the time step  $\Delta t$ is small on the time scale of the damping) per memory time, $N=T_{\rm m}/\Delta t$, is not too large, as keeping track of the system state a long time in the past is computationally demanding.

\section{Quantum trajectories from a measurement perspective}
In the rotating wave approximation the coupling of a two level atom to an electromagnetic field can be described by the interaction picture Hamiltonian 
\begin{equation}
H_{I}(t)=i\hbar[\xi^{\dagger}(t)\sigma_{I}(t)-\xi(t)\sigma^{\dagger}_{I}(t)],
\end{equation}
where $\sigma_{I}(t)$ is the interaction picture atomic lowering operator  and 
\begin{equation}
\xi(t)=\sum_{k}g_{k}b_{k}(t_{0})e^{-i\omega_{k}(t-t_{0})}
\end{equation}
is referred to as the driving field \cite{qn}.  The $k$th mode of the field $b_{k}$, where $[b_{k},b^{\dagger}_{k'}]=\delta_{kk'}$, is coupled to the atom with a strength $g_{k}$ and has a frequency of oscillation $\omega_{k}$.

Quantum trajectories can be derived from a measurement theory perspective by considering continuous measurements made on the electromagnetic field that has  been  emitted irretrievably from the atom \cite{jack}.  The probability of getting a set of results $R(t,t_{0})$  over the interval $[t_{0},t]$ given a set of inputs $I(t,t_{0})$ is, 
\begin{equation}
P_{\rm RI}(t,t_{0})=\langle\psi(t_{0})|\Omega_{\rm RI}^{\dagger}(t,t_{0})\Omega_{\rm RI}(t,t_{0})|\psi(t_{0})\rangle,
\end{equation}
where 
\begin{equation}
\Omega_{\rm RI}(t,t_{0})=\langle {\rm R}(t,t_{0})|U_{\rm int}(t,t_{0})|{\rm I}(t,t_{0})\rangle
\end{equation}
is the evolution operator for the atomic system conditioned on the particular results and  inputs, and $U_{\rm int}(t,t_{0})$ is the unitary evolution operator of the total system of atom and reservoir.

Consider a process in discrete time where the time between any two points on our grid is $\Delta t$. We can simulate a continuous measurement process by determining the conditional probability of getting a particular result at time $t$ given a set of previous results. This conditional probability can be written as
\begin{equation}
\frac{P_{\rm RI}(t,t_{0})}{P_{\rm RI}(t-\Delta t,t_{0})}=\frac{\langle \tilde{\psi}^{0}(t)|\tilde{\psi}^{0}(t)\rangle}{\langle\tilde{\psi}^{0}(t-\Delta t)|\tilde{\psi}^{0}(t-\Delta t)\rangle},\label{conditional}
\end{equation}
where the unnormalized wave function satisfies the interaction picture equation
\begin{equation}
|\tilde{\psi}^{0}(t)\rangle=|\tilde{\psi}^{0}(t-\Delta t)\rangle+\Delta\Omega_{\rm IR}(t,t-T_{\rm m})\circ |\tilde{\psi}(t-T_{\rm m})\rangle,\label{general}
\end{equation}
where we have assumed that there is a finite time in the past, $T_{\rm m}$, before which the state of the system does not effect the probabilities in the present. We will discuss this in more detail below and also illuminate the meaning of the superscript. 

The explicit form of (\ref{general}) for a photon counting measurement situation with vacuum input (where the time step is chosen so that the probability of two particles being detected during one time step is negligible) is
 \begin{eqnarray}
\fl |\tilde{\psi}^{0}(t)\rangle  =  |\tilde{\psi}^{0}(t-\Delta t)\rangle - \Delta t\sigma_{I}(t)^{\dagger}\int^{t}_{t-T_{\rm m}}\!\!\!\!ds\,f_{\rm 
        m}(t-s)T\left\{\sigma_{I}(s)\prod_{j}{\mathcal A}(t_{j})V_{0}\right\}\circ|\tilde{\psi}(t-T_{\rm m})\rangle
\nonumber \\
\fl + \Delta N(t)\left[\int^{t}_{t-T_{\rm m}}
                \!\!\!\! ds\,h(t-s)T\left\{\sigma_{I}(s)\prod_{j}{\mathcal A}(t_{j}) V_{0}\right\}\circ|\tilde{\psi}(t-T_{\rm m})\rangle- |\tilde{\psi}^{0}(t-\Delta t)\rangle \right] ,\label{equation}
\end{eqnarray} 
where the interaction picture operators inside the curly brackets, $T\{\cdots\}$, are time-ordered. The restriction $\Delta N(t)^{2}=\Delta N(t)$ 
 describes a point process in which at time $t$ either  
one photon is detected, $\Delta N(t)=1$, or 
no photons are detected, $\Delta N(t)=0$.
 The $t_{j}$ are the discrete times of previous detections during the time interval $[t-T_{\rm m},t)$ and
\begin{equation}
{\mathcal A}(t_{j})\equiv\int^{t_{j}}_{t-T_{\rm m}}\!\!\!\! ds\, h(t_{j}-s)\sigma_{I}(s),
\end{equation}
where 
\begin{equation}
h(t-s)=\sum_{k} g_{k}e^{-i\omega_{k}(t-s)},
\end{equation}
is the single particle Green's function for a photon propagating from the atom to the detector. We call this quantity the {\it response} as it represents the weighting of a particular photon emission time given an instantaneous measurement at time $t$.
The interaction picture evolution operator has been reduced to the form
\begin{equation}
 V_{0}=\exp\left\{-\int^{t}_{t-T_{\rm 
m}}\!\!\!\! ds_{1}\int^{s_{1}}_{t-T_{\rm m}} \!\!\!\!ds_{2}\, \sigma_{I}^{\dagger}(s_{1})f_{\rm m}(s_{1}-s_{2})\sigma_{I}(s_{2})\right\},
\end{equation}
where
\begin{equation}
f_{\rm m}(t-s)=\sum_{k} |g_{k}|^{2}e^{-i\omega_{k}(t-s)},
\end{equation}
is the single particle Green's function for a photon that is emitted and then reabsorbed again by the system. This we call the {\it memory function} of the system \cite{qn}. 

A trajectory is simulated by calculating the probability of a click of the detector $\langle\tilde{\psi}^{0}(t)|\tilde{\psi}^{0}(t)\rangle_{\Delta N(t)=1}\times\Delta t$ (or no click of the detector $\langle\tilde{\psi}^{0}(t)|\tilde{\psi}^{0}(t)\rangle_{\Delta N(t)=0}$), dividing by $\langle\tilde{\psi}^{0}(t-\Delta t)|\tilde{\psi}^{0}(t-\Delta t)\rangle$ and comparing this conditional probability with a random number between 0 and 1. If the probability is greater (less) than the random number then we say that a photon  has been irretrievably emitted from the system at time $t$.  This result becomes part of the permanent measurement record and is incorporated into the subsequent evolution as an additional factor, ${\mathcal A}(t)$. 
The rate of fall off of the response and the memory function determines the time $t-T_{\rm m}$, where $T_{\rm m}$ is referred to as the memory time. We can neglect the effect of earlier emissions on the probabilities at time $t$. The state of the system during the memory time is difficult to define because subsequent measurements could alter the state. However, the state before time $t-T_{\rm m}$ will not change with future measurements so we can fix a  state at time $t-T_{\rm m}$ conditioned on measurements up until time $t$. It is only in the Markovian case that the conditioned evolution operator $\Omega_{RI}$ factorizes at time $t-T_{\rm m}$, as in this case the memory function decays to zero during a single time step and we can put $T_{\rm m}=\Delta t$. In general, the entanglement between the system and bath at the time $t-T_{\rm m}$ means that the  conditioned state of the reduced system is mixed.  The $\circ$-product of (\ref{general}) represents a summation over  separate conditioned evolution operators acting on separate kets in the system Hilbert space,
\begin{equation}
\Delta\Omega_{\rm IR}(t,t-T_{\rm m})\circ|\tilde{\psi}(t-T_{\rm m})=\sum_{l}\Delta\Omega^{l}_{\rm IR}(t,t-T_{\rm m})|\tilde{\psi}^{l}(t-T_{\rm m})\rangle.
\end{equation}
 The kets of the system with different superscript labels are orthogonal as the labels represent a different state of the field. The conditioned state of the system is a density matrix given by
\begin{equation}
 \rho_{c}(t-T_{\rm m})= \frac{{\displaystyle\sum}_{l}|\tilde{\psi}^{l}(t-T_{\rm m})\rangle\langle\tilde{\psi}^{l}(t-T_{\rm m})|}{{\displaystyle\sum}_{l}\langle\tilde{\psi}^{l}(t-T_{\rm m})|\tilde{\psi}^{l}(t-T_{\rm m})\rangle}\label{state}
\end{equation}
and conditioned expectation values of system quantities are then given by $\langle \hat{A}(t)\rangle_{c}={\rm Tr}\{\rho_{c}(t)\hat{A}\}$.

\section{Simulating non-Markovian quantum trajectoies}
In the above section (and in our previous work \cite{jack}) we have given a formal derivation of non-Markovian trajectories. The present task is to determine an algorithm to simulate a trajectory numerically. Equation (\ref{equation}) clarifies the connection between a measurement event and the processes of emission and absorption going on in the atom.  At time $t$ there is a chance of a photon being detected which corresponds to a photon being irretrievably emitted at some time in the past. The times of emission are weighted by the  response function (see the third term on the RHS of  (\ref{equation})). There is also the chance of a photon not being detected which corresponds to a photon being emitted at some time in the past and then reabsorbed by the system. The emission times are weighted by the memory function (see the second term on the RHS of  (\ref{equation})). 

In a simulation, the measurement and the emission events take place at discrete times on a grid and the number of time steps per memory time is given by $N=T_{\rm m}/\Delta t$. In figure \ref{traj} we have given a diagrammatic representation of the three simplest cases. To keep track of whether a photon corresponding to a measurement event has or has not been emitted at every discrete time  we introduce a notation consisting of a binary number associated with each possibility where a $1$ denotes that a photon has already been  emitted and a $0$ the opposite. Each place in the binary number corresponds to a different measurement event. As an example, we consider the state of the system at time $t-T_{\rm m}$ in the  $N=3$ case. A schematic of the situation is given in figure \ref{shade}.  There are four possible paths that the system can take at time $t-2\Delta t$ and a ket is associated with each of these possible paths. This example  demonstrates a general property of these trajectories, in that, if there are $N$ grid points in a memory time and each of these points corresponds to a measurement event then there are $2^{N-1}$ possible paths for the system to take at the time $t-T_{\rm m}$. One therefore needs $2^{N-1}$ kets to fully specify the conditioned state. The rapid growth of this quantity with increasing $N$ puts a fundamental limitation on this method as a ket must be stored for each of these possibilities.  Although the kets are in the Hilbert space of the system the different labels correspond to orthogonal states of the reservoir because they either correspond to distinct measurement results or to different numbers of photons in the field and therefore the complete state is an incoherent sum of the kets and is given by  (\ref{state}).

The above  notation is also useful for determining an algorithm to compute the forward propagation of the kets. 
As an example we consider an explicit calculation of the measurement probabilities at time $t$ from the state at $t-3\Delta t$ for the $N=3$ case. We define kets of the form $|\tilde{\psi}^{lmn}(t)\rangle$ where the right superscript label corresponds to the measurement event at $t-2\Delta t$ (with a corresponding emission weighting function $f_{1}=h$) the middle superscript to the measurement event at $t-\Delta t$ (with associated function $f_{2}=f_{\rm m}$) and the left superscript to a possible event at time $t$ (with function $f_{3}=h$ or $f_{\rm m}$), see figure \ref{traj}. A first-order algorithm to calculate numerically  the first increment of the kets in the Schr\"{o}dinger picture version of  (\ref{equation}) is
\begin{eqnarray}
\eqalign{ |\tilde{\psi}^{000}(t-2\Delta t)\rangle  = (1-\rmi H_{0}\Delta t)|\tilde{\psi}^{00}(t-3\Delta t)\rangle,\label{algorithim1}\\
 |\tilde{\psi}^{001}(t-2\Delta t)\rangle  = (1-\rmi H_{0}\Delta t)|\tilde{\psi}^{01}(t-3\Delta t)\rangle+ \sigma\overline{f_{1}}(0)|\tilde{\psi}^{00}(t-3\Delta t)\rangle,\label{algorithim2}\\
{\vdots} \\
\fl |\tilde{\psi}^{011}(t-2\Delta t)\rangle  = (1-\rmi H_{0}\Delta t)|\tilde{\psi}^{11}(t-3\Delta t)\rangle\\
+\sigma\left(\overline{f_{2}}(\Delta t)|\tilde{\psi}^{01}(t-3\Delta t)\rangle+ \overline{f_{1}}(0)|\tilde{\psi}^{10}(t-3\Delta t)\rangle\right),\\
{\vdots}}
\end{eqnarray}
where we have considered the averaged functions
\begin{equation}
\overline{f_{j}}(t_{j}-t)=\int^{t}_{t-\Delta t}ds f_{j}(t_{j}-s)
\end{equation}
 as the rate of change of the memory function may be faster than the time scale of interest. 
We must now rearrange our kets before the next time step as we have reached the $t-2\Delta t$ measurement event. We therefore keep only  the kets that are labeled with a right superscript of $1$ as these represent the paths where an emission corresponding to the measurement result at $t-2\Delta t$ has occurred. The other states become redundant.  The new states are given by 
\begin{equation}
\fl|\tilde{\psi}^{001}(t-2\Delta t)\rangle\rightarrow|\tilde{\psi}^{00}(t-2\Delta t)\rangle,\qquad|\tilde{\psi}^{011}(t-2\Delta t)\rangle\rightarrow |\tilde{\psi}^{01}(t-2\Delta t)\rangle,\qquad\cdots
\end{equation}
and
$f_{2}\rightarrow f_{1}$ and $f_{3}\rightarrow f_{2}$. The next increment  is then
\begin{equation}
\eqalign{|\tilde{\psi}^{00}(t-\Delta t)\rangle = (1-\rmi H_{0}\Delta t)|\tilde{\psi}^{00}(t-2\Delta t)\rangle,\\
 |\tilde{\psi}^{01}(t-\Delta t)\rangle  = (1-\rmi H_{0}\Delta t)|\tilde{\psi}^{01}(t-2\Delta t)\rangle+\sigma \overline{f_{1}}(0)|\tilde{\psi}^{00}(t-2\Delta t)\rangle,\\
 |\tilde{\psi}^{10}(t-\Delta t)\rangle  =(1-\rmi H_{0}\Delta t)|\tilde{\psi}^{10}(t-2\Delta t)\rangle+ \sigma\overline{f_{2}}(\Delta t)|\tilde{\psi}^{00}(t-2\Delta t)\rangle,\\
\fl|\tilde{\psi}^{11}(t-\Delta t)\rangle  = (1-\rmi H_{0}\Delta t)|\tilde{\psi}^{11}(t-2\Delta t)\rangle\\
+ \sigma\left(\overline{f_{1}}(0)|\tilde{\psi}^{10}(t-2\Delta t)\rangle+ \overline{f_{2}}(\Delta t)|\tilde{\psi}^{01}(t-2\Delta t)\rangle\right).}
\end{equation}
This time the measurement event corresponds to no detection so rearranging the states must proceed by 
\begin{equation}
\eqalign{|\tilde{\psi}^{0}(t-\Delta t)\rangle  = |\tilde{\psi}^{00}(t-\Delta t)\rangle+\sigma^{\dagger}|\tilde{\psi}^{01}(t-\Delta t)\rangle,\\
|\tilde{\psi}^{1}(t-\Delta t)\rangle =|\tilde{\psi}^{10}(t-\Delta t)\rangle+\sigma^{\dagger}|\tilde{\psi}^{11}(t-\Delta t)\rangle,}
\end{equation}
and $f_{2}\rightarrow f_{1}$. The final increment is
\begin{equation}
\eqalign{|\tilde{\psi}^{0}(t)\rangle  = (1-\rmi H_{0}\Delta t)|\tilde{\psi}^{0}(t-\Delta t)\rangle,\\
|\tilde{\psi}^{1}(t)\rangle  = (1-\rmi H_{0}\Delta t)|\tilde{\psi}^{1}(t-\Delta t)\rangle+\sigma\overline{f_{1}}(0)|\tilde{\psi}^{0}(t-\Delta t)\rangle.}
\end{equation}
The rearrangement at the end to get a final single ket $|\tilde{\psi}^{0}(t)\rangle$ will obviously depend on whether $f_{1}$ corresponds to a detection or no detection.
For example, if $f_{1}=f_{\rm m}$ then $|\tilde{\psi}^{0}(t)\rangle=|\tilde{\psi}^{0}(t)\rangle+\sigma^{\dagger}|\tilde{\psi}^{1}(t)\rangle$ and the conditional probability of no detection at time $t$ given the previous measurement record is given by  (\ref{conditional})
where $|\tilde{\psi}^{0}(t-\Delta t)\rangle$ is the ket corresponding to the realization of the simulated measurement at time $t-\Delta t$. 
We have written out the explicit procedure for the $N=3$ case but the basic structure of the general case should be apparent from the above example. An increment in each ket has the form
\begin{equation}
|\tilde{\psi}^{l}(t+\Delta t)\rangle=(1-\rmi H_{0}\Delta t)|\tilde{\psi}^{l}(t+\Delta t)\rangle+\sigma\sum_{n}{\overline f}_{n}|\tilde{\psi}^{n}(t)\rangle,
\end{equation}
where the $f_{n}$ are taken from the set of responses and memory functions corresponding to the measurement record and are multiplied by the appropriate ket, $|\tilde{\psi}^{n}(t)\rangle$. The summation is over all the possible paths to  $|\tilde{\psi}^{l}(t+\Delta t)\rangle$. The binary notation can be exploited to make this step automatic. The kets are then reduced by referring to the measurement record.

In summary, the system state is propagated forward by following all the possible paths that the system can take constrained by the fact that the paths must be consistent with the measurement record. The measurement probabilities at time $t$ are determined by propagating the system forward from time $t-T_{\rm m}$ conditioned on the previous measurement record. The time $t-T_{\rm m}$ is chosen because at this is the latest time where the state of the system is independent of the outcome of the measurement at time $t$. The outcome of the (simulated) measurement at time $t$ becomes part of the permanent measurement record  and the system state at time $t-T_{\rm m}$ can be propagated  forward $\Delta t$ in accordance with this record.  $2^{N-1}$  kets must be stored to fully represent the conditioned state of the system at any time. By exploiting a special binary notation an algorithm can be determined that takes any memory function and response and the associated memory time and generates a trajectory.  The Markovian trajectory is then a special case of this algorithm when $N=1$.

\section{An atom radiating into a cavity}
The practicality of the method is demonstrated by comparing the results of a non-Markovian trajectory with the results of a traditional Markovian trajectory for an extended system, using the fact that systems that undergoes non-Markovian decay can often be embeded in a larger system that undergo Markovian decay \cite{imamoglu,garraway}. 

 The simple test case we consider here is a driven two level atom coupled to a single mode of a one-sided optical cavity, where we assume that the atom only emits into the cavity. From a non-Markovian perspective the system is the two level atom given by the Hamiltonian (in a frame rotating with the classical driving field)
\begin{equation}
H_{0}=\Delta\omega\sigma_{z}+\frac{\Omega}{2}(\sigma+\sigma^{\dagger}),
\end{equation}
where $\Delta\omega$ is the detuning from the driving field and $\Omega$ is the strength of the driving field.  The memory and response functions are given in terms of Fourier transforms by
\begin{eqnarray}
f_{\rm m}(t-s)&=&\frac{\gamma}{\sqrt{2\pi}}\int d\omega \frac{\kappa}{(\frac{\kappa}{2})^{2}+(\omega-\nu)^{2}}e^{-\rmi\omega(t-s)},\\
h(t-s)&=&\sqrt{\frac{\gamma}{2\pi}}\int d\omega \frac{\sqrt{\kappa}}{\frac{\kappa}{2}+i(\omega-\nu)}e^{-\rmi\omega(t-s)},
\end{eqnarray}
where $\gamma$ is the strength of the coupling to the atom, $\nu$ is the frequency of the cavity in the rotating frame and $\kappa$ is the line-width of the cavity.
On the other hand by modeling the cavity explicitly one could run a Markovian simulation with the effective Hamiltonian
\begin{equation}
H_{\rm eff}=H_{0}+\nu a^{\dagger}a+\rmi\sqrt{\gamma}(a\sigma^{\dagger}-a^{\dagger}\sigma)-\rmi\frac{\kappa}{2}a^{\dagger}a,
\end{equation}
where $a$ is the annihilation operator of the cavity mode and the collapse operator is $\sqrt{\gamma}a$. A comparison of the results of these two systems are shown in figure \ref{results} and \ref{waittime}.
 The non-Markovian quantum trajectory determines the state of the atom conditioned on measurements of the emitted light.  By specifying the state of the atom after all the light that the atom has emitted has been measured we gain maximum knowledge about the atom. 
In contrast, in the extended system method the state of the atom, calculated by tracing the cavity out of the conditioned state of atom and cavity, implies a partial erasing of information about the state  of the atom. 
The ensemble average of these two methods is equivalent as the average sums the measurement results over all times.
In figure \ref{single1} and \ref{single2} we compare a single quantum trajectory with an extended system trajectory.

\section{Summary}
We have presented a generalization of the quantum trajectory method for treating an atomic system radiating into a reservoir to the non-Markovian regime.   We have summarized the derivation of these trajectories from a continuous  measurement perspective and explicitly demonstrated the algorithm for numerically simulating a non-Markovian quantum trajectory.
The results of numerical simulations of a simple test example were shown to agree with more traditional methods. The method retains many of the familiar concepts of Markovian quantum trajectories and, in fact, contains the Markovian method as a special case. In the Markovian case the reservoir enters the equations of motion for the reduced system as a single rate which is determined by integrating the memory and response functions over all time. The non-Markovian case requires a more detailed description of the reservoir and the time dependence of the memory and response functions  becomes important.

The method described here promises to have wide applicability to many situations where weakly non-Markovian behavior (i.e., not too many time steps during a memory time)  arises.

\ack
The authors are grateful for the support of the Marsden Fund of the Royal Society of New Zealand and the University of Auckland Research Fund.

\Figures
 \Figure{\protect\label{traj}Schematic of a non-Markovian trajectory. The grid on the x-axis represents the propagation of discrete time from left to right. The curves ending on the axis represent photons being emitted and then reabsorbed by the system, whereas, the straight lines ending above the axis represent the irretrievable emission of a photon from the system. The measurement record fixes the events at the end of the lines but photons can be emitted at any time after the beginning of the lines up until the lines end. The dotted lines ending at $t$ represent the two possible outcomes of a measurement at this time. The situation depicted here is: at time $t-\Delta t$ no photon was detected and at $t-2\Delta t$ one photon was detected.}
\Figure{\protect\label{shade}Diagrammatic representation of the four possible combinations of emission times that contribute to the two measurement results at $t-\Delta t$ and $t-2\Delta t$ of figure \ref{traj}, $N=3$. The possible regions of emission  are represented by the shading. The darker shading represents an overlap of the two emission regions. The possibilities are: $00$ - both the  photon detected at $t-2\Delta t$ and the photon reabsorbed at time $t-\Delta t$ have not been emitted, $10$ - only the  photon detected at $t-2\Delta t$ has been emitted, $01$ - only the  photon reabsorbed at $t-\Delta t$ has been emitted, $11$ - both the  photon detected at $t-2\Delta t$ and the photon reabsorbed at time $t-\Delta t$ have been emitted.}
\Figure{\protect\label{results} The real (upper) and imaginary (lower) parts of the correlation function of the system $\langle \sigma^{\dagger}(\tau)\sigma\rangle$ determined by an average over a single trajectory of $10^{5}$ detections is plotted in (a) for the parameters $\Omega=2\gamma$, $\nu=2\gamma$, $\Delta \omega=0$ and $\kappa=10\gamma$. The spectrum (the Fourier transform of the correlation function) is plotted in (b). Note the asymmetry of the spectrum due to the cavity being detuned $\nu\neq 0$ from the central peak of the Mollow spectrum. The corresponding quantities from the extended system treatment are also given (dashed line). For comparison we have also plotted the Mollow spectrum for an atom radiating into free space (dotted line).}
\Figure{\protect\label{waittime} The waiting time distribution for the detections is compared with that calculated by the extended system treatment (dashed line).}
\Figure{\protect\label{single1} A typical run of ten detections of a single non-Markovian quantum trajectory. We have plotted the probability of a detection as a function of time for the parameters: $\kappa=8\gamma$, $\Omega=2\gamma$ and $\Delta\omega=\nu=0$. The same probability determined from the extended system method using the same random numbers is also shown (dashed line).}
\Figure{\protect\label{single2} Plot of the evolution of the conditioned expectation value of $\sigma_{z}$ over the trajectory. The extended system method (ESM) and the non-Markovian quantum trajectory (NMQT) calculate different states for the atom. Notice that in the NMQT case the atom makes full oscillations and also starts emitting before a detection occurs.}

\newpage

\begin{figure} 
{\bf \Large Figure.1.}
\vspace{1cm}
\begin{center}
\includegraphics[height=0.25\linewidth,width=0.6\linewidth]{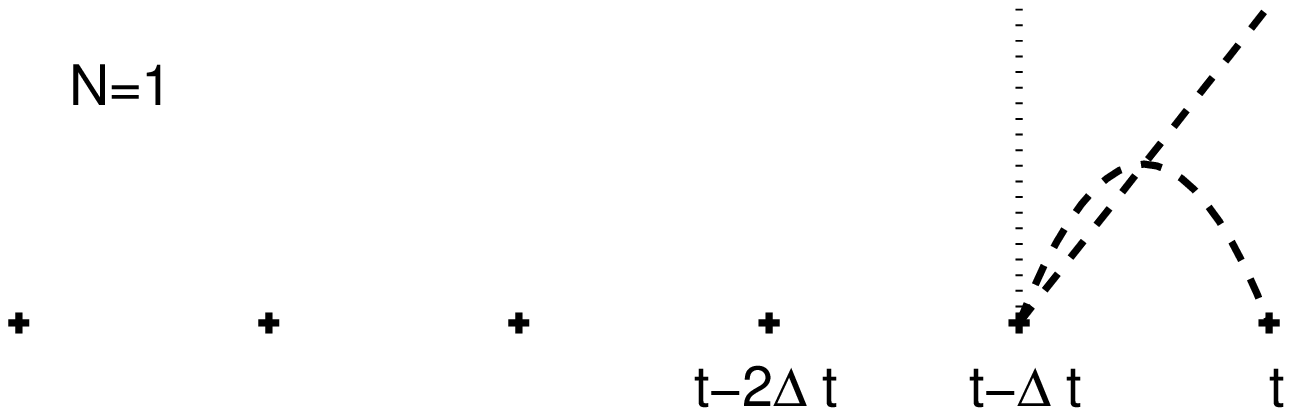}
\end{center}
\begin{center}
\includegraphics[height=0.25\linewidth,width=0.6\linewidth]{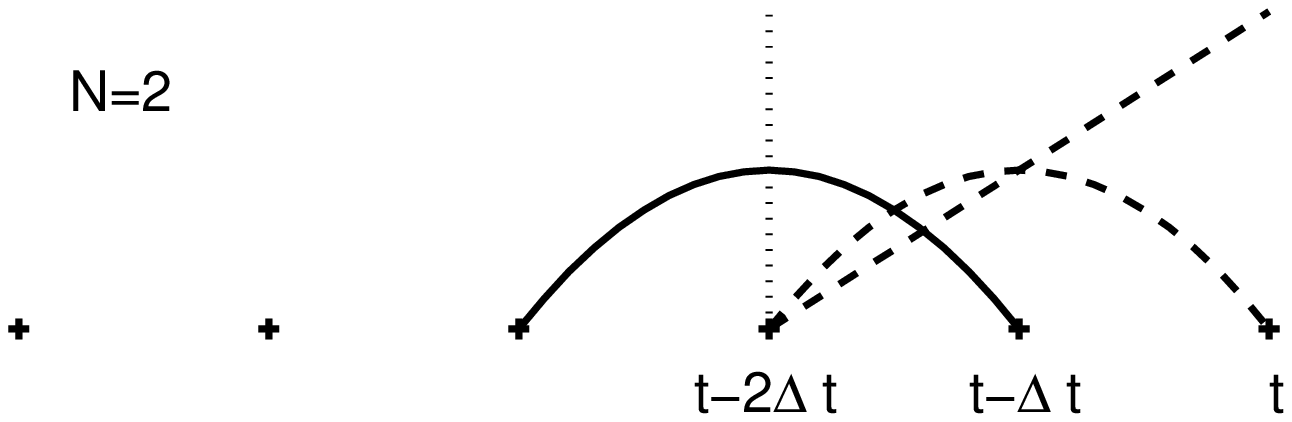}
\end{center}
\begin{center}
\includegraphics[height=0.25\linewidth,width=0.6\linewidth]{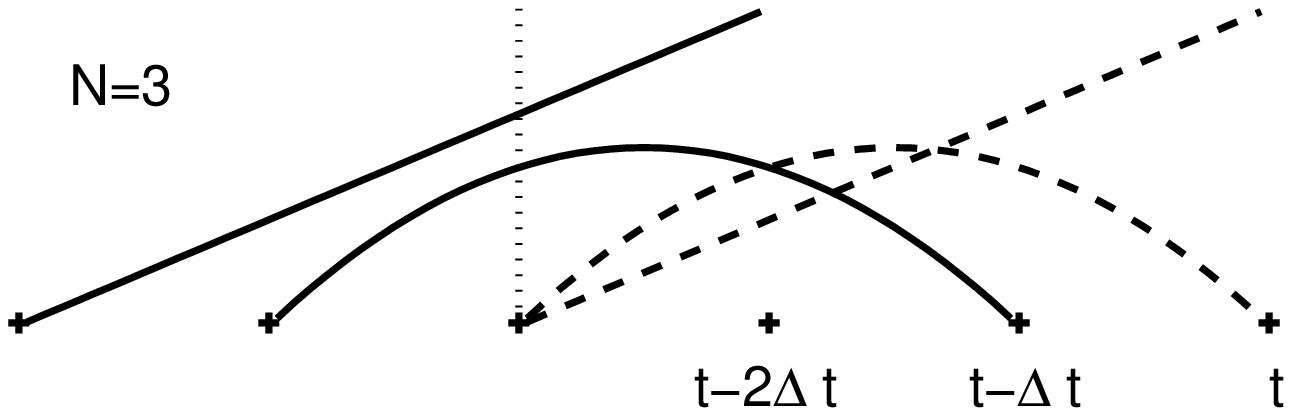}

\end{center}
\end{figure} 
\newpage

\begin{figure}
{\bf \Large Figure.2.}
\vspace{1cm}
\begin{center}
\includegraphics[height=0.6\linewidth,width=0.7\linewidth]{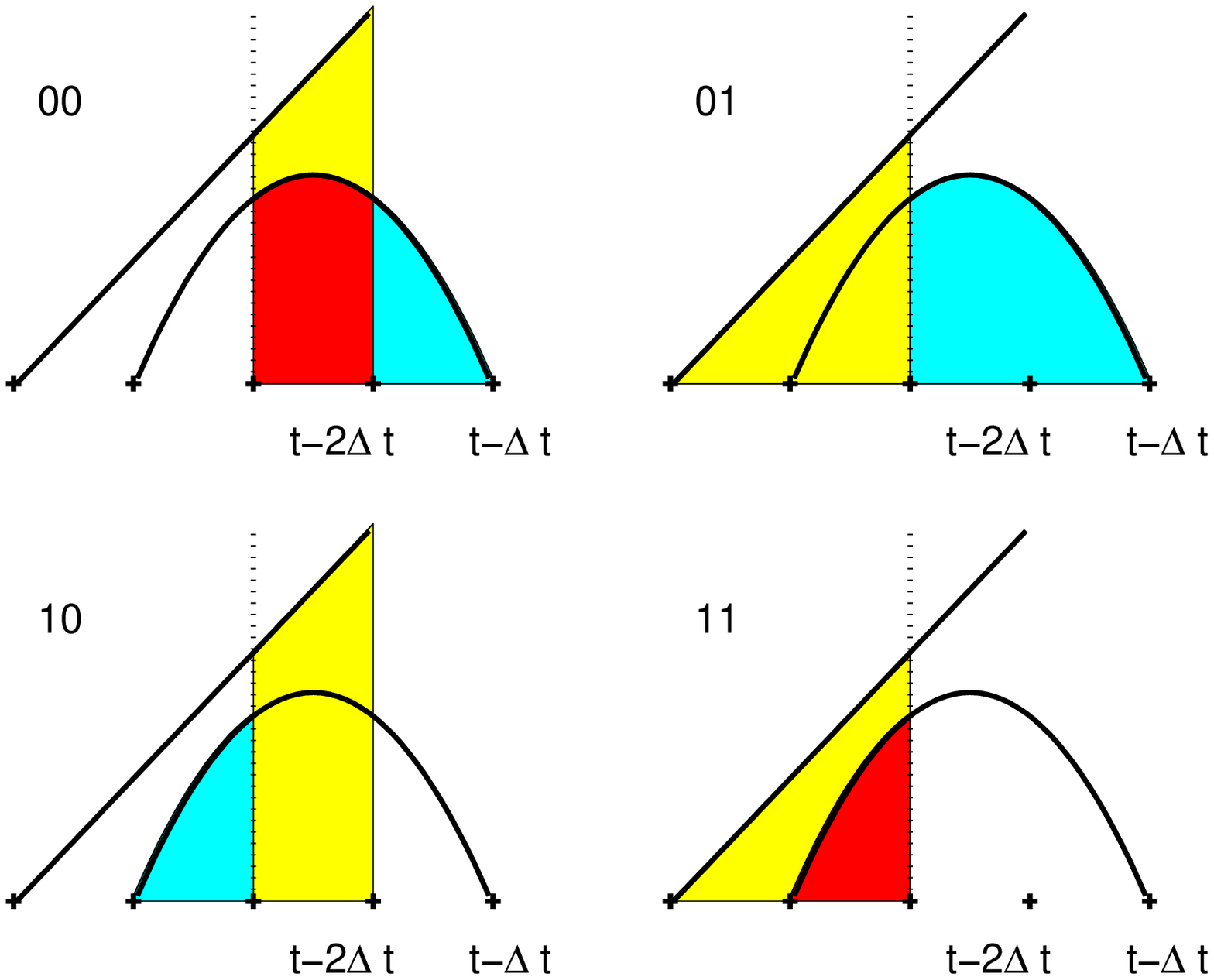}

\end{center}

\end{figure}
 
\newpage
\begin{figure}
{\bf \Large Figure.3.(a)}

\vspace{1cm}
\begin{center}
\includegraphics[height=0.6\linewidth,width=0.7\linewidth]{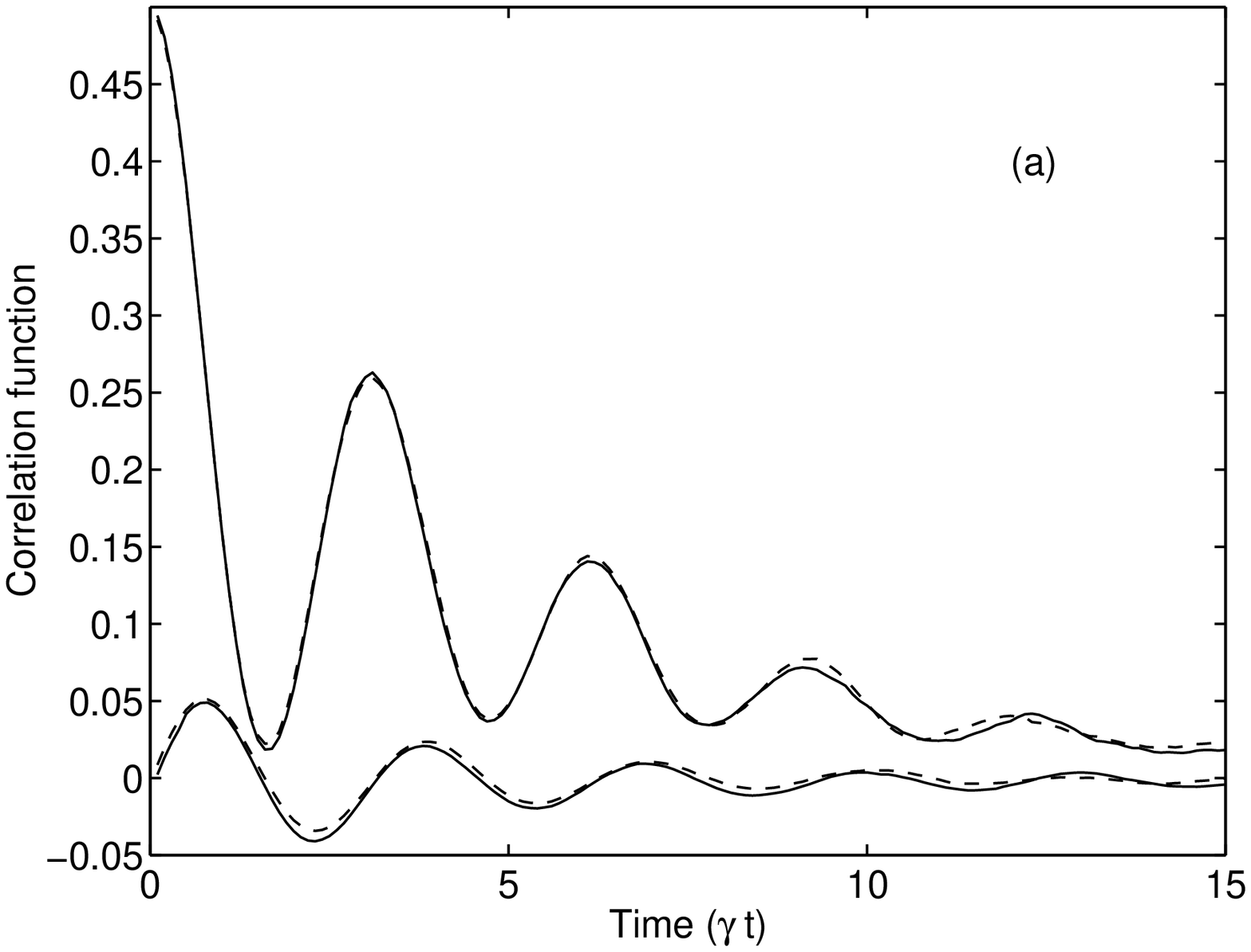}
\end{center}
\end{figure}
\newpage
\begin{figure}
{\bf \Large Figure.3.(b)}
\vspace{1cm}
\begin{center}
\includegraphics[height=0.6\linewidth,width=0.7\linewidth]{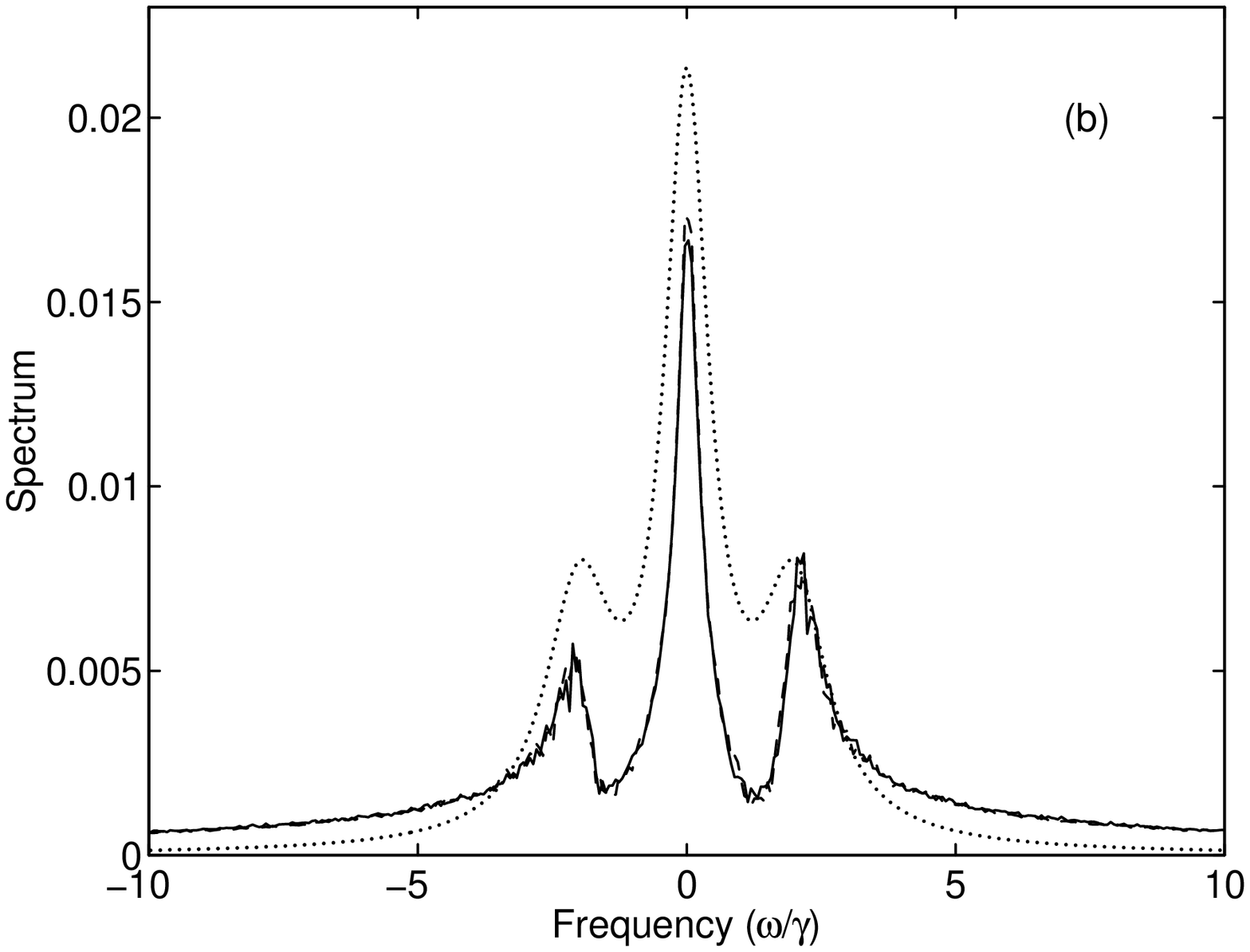}
\end{center}
\end{figure}

\newpage

\begin{figure}
{\bf \Large Figure.4.}
\vspace{1cm}
\begin{center}
\includegraphics[height=0.6\linewidth,width=0.7\linewidth]{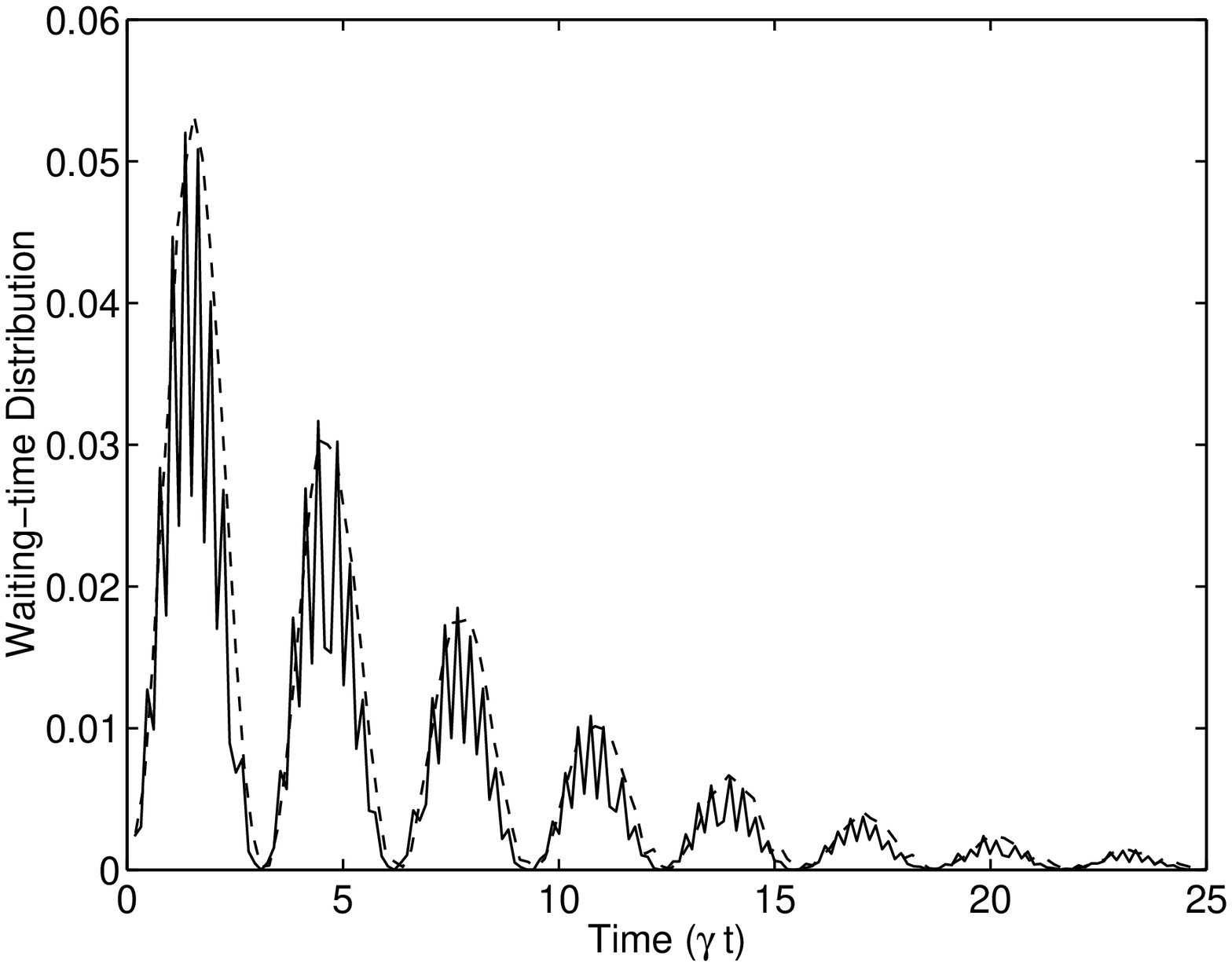}
\end{center}

\end{figure}
\newpage

\begin{figure}
{\bf \Large Figure.5.}
\vspace{1cm}
\begin{center}
\includegraphics[height=0.5\linewidth,width=0.7\linewidth]{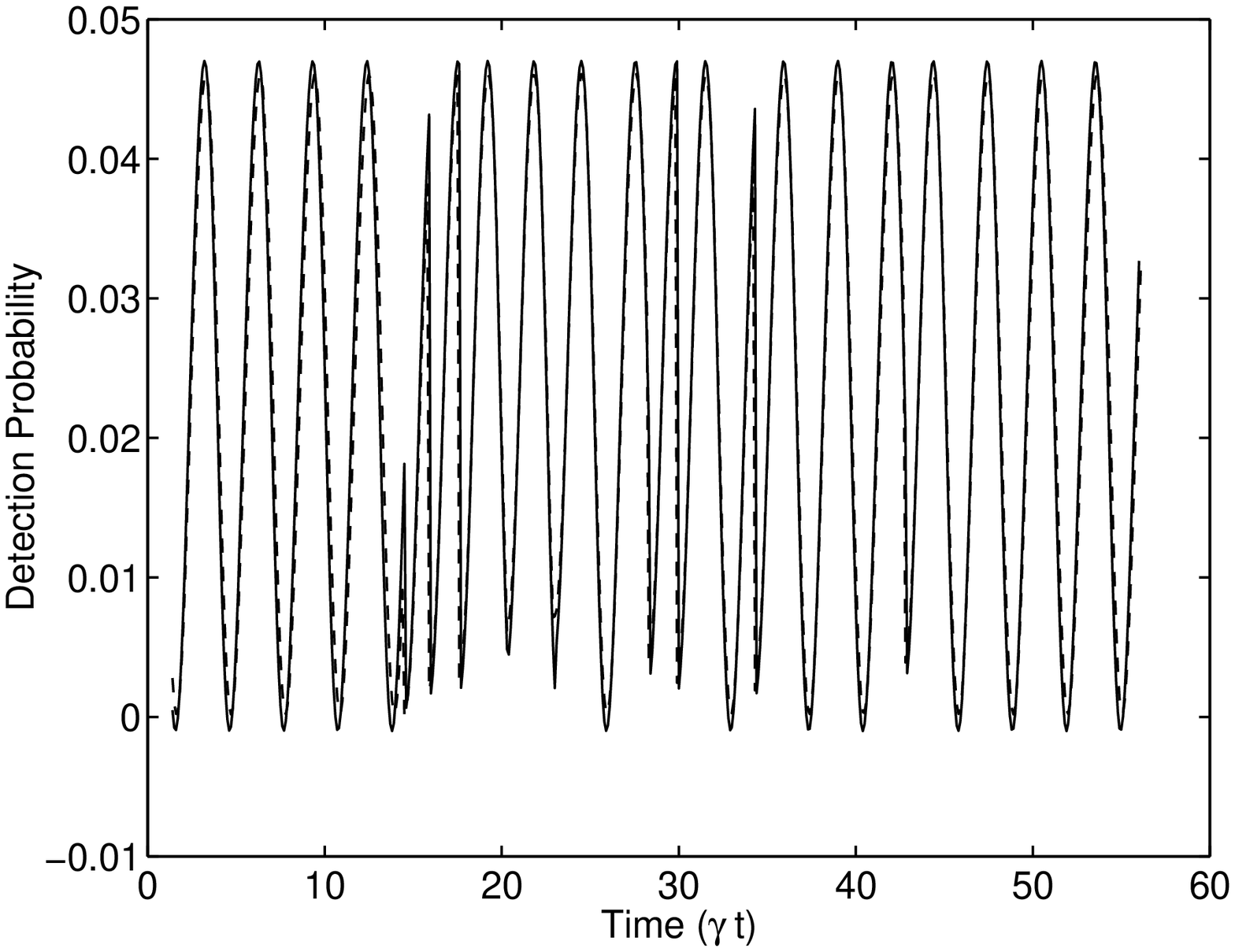}
\end{center}

\end{figure}

\newpage

\begin{figure}
{\bf \Large Figure.6.}
\vspace{1cm}
\begin{center}
\includegraphics[height=0.5\linewidth,width=0.7\linewidth]{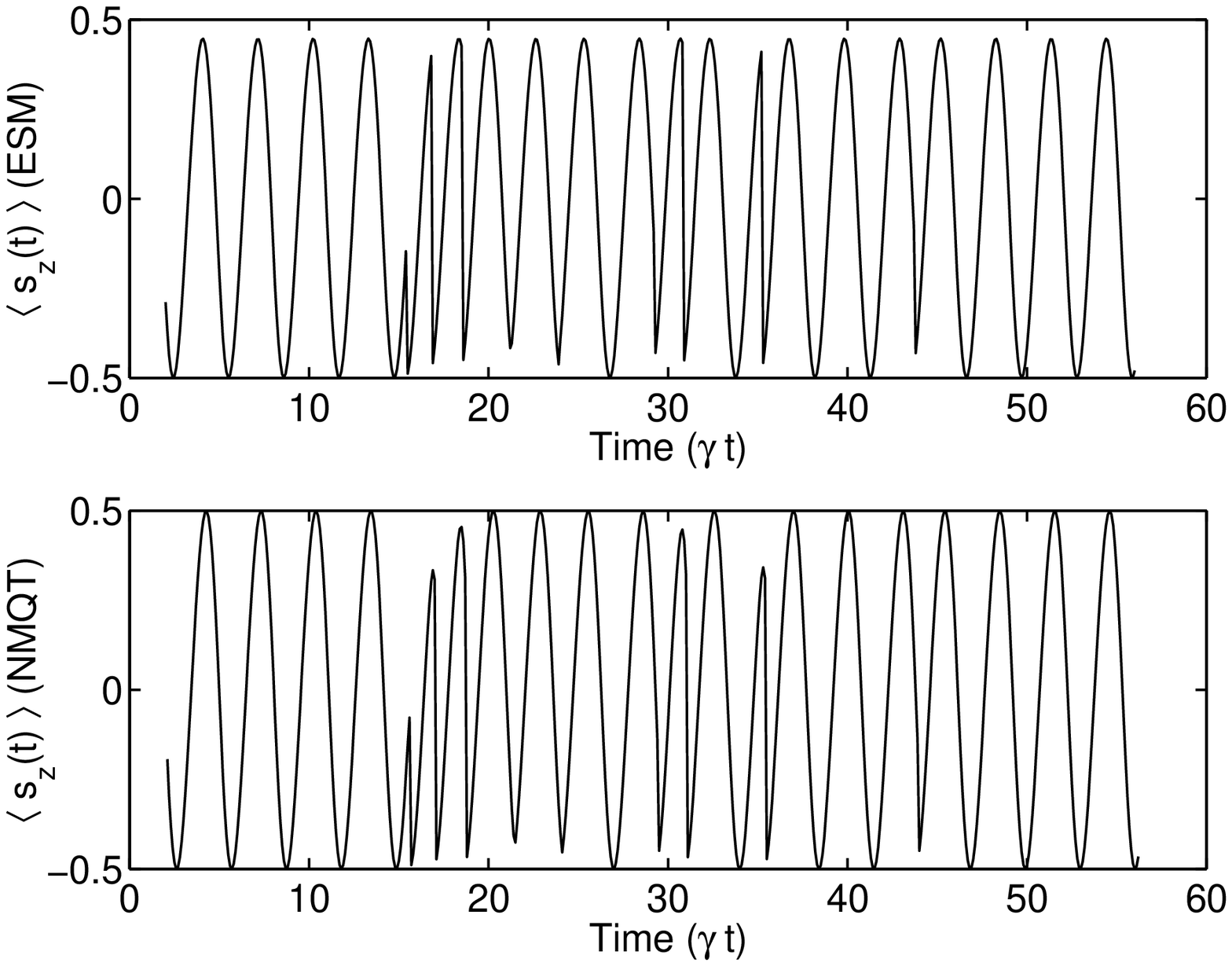}
\end{center}

\end{figure}
\end{document}